\documentclass[aps,prd,twocolumn,showpacs,preprintnumbers,nofootinbib,amsmath,amssymb]{revtex4}

\usepackage{graphicx}
\usepackage[caption=false]{subfig}
\usepackage{bm}
\usepackage{amsthm,url}
\usepackage{color}

\def\be{\begin{equation}}
\def\ee{\end{equation}}
\def\bea{\begin{eqnarray}}
\def\eea{\end{eqnarray}}
\def\bn{\begin{enumerate}}
\def\en{\end{enumerate}}
\def\ll{\left}
\def\rr{\right}

\def\S{\mathcal{S}}
\def\k{\mathbf{k}}
\def\K{\mathcal{K}}
\def\U{\mathbf{U}}
\def\u{\mathbf{u}}

\def\d{\mathbf{d}}

\def\mb{\mathbf}

\def\mb{\mathbf}
\def\h{\mb{h}}
\def\th{\tilde{\mb{h}}}

\def\d{\mb{d}}
\def\e{\mb{e}}
\def\n{\mb{n}}
\def\N{\mathcal{N}}

\def\F{\text{\bf F}}
\def\D{\mb{D}}

\def\U{\mb{U}}
\def\u{\mb{u}}
\def\hu{\hat{\mb{u}}}

\def\V{\mb{V}}

\def\z{\mb{z}}

\def\X{\mathcal{X}}
\def\x{\mb{x}}
\def\s{\mb{s}}

\def\cA{\mathcal{A}}
\def\n{\mb{n}}

\usepackage{color}

\makeatletter

\newcommand{\Rmnum}[1]{\expandafter\@slowromancap\romannumeral #1@}
\makeatother

\begin{document}

\preprint{------}

\title{Synthetic streams in a Gravitational Wave inspiral search with a multi-detector network} 
\author{Haris K\footnote{Electronic address: haris@iisertvm.ac.in} and Archana Pai\footnote{Electronic address: archana@iisertvm.ac.in}}
\affiliation{Indian Institute of Science Education and Research Thiruvananthapuram, CET Campus, Trivandrum 695016}
\date{\today}
\begin{abstract}
Gravitational Wave Inspiral search with a global network of interferometers when carried in a phase coherent fashion would mimic an
{\it effective} multi-detector network with synthetic streams constructed by the linear combination of the data from different detectors.
For the first time, we demonstrate that the two synthetic data streams pertaining to the two polarizations of Gravitational Wave can be derived prior to the
maximum-likelihood analysis in a most natural way using the technique of singular-value-decomposition applied to the network
signal-to-noise ratio vector. We construct the {\it network matched filters} in combined network plus spectral space
which capture both the synthetic streams. We further show that the network LLR is then sum of the LLR of each synthetic stream. The four
extrinsic parameters namely $\{A_0,\phi_a,\epsilon,\psi\}$ are mapped to the two amplitudes and two phases namely $\{\boldsymbol{\rho}_L,\boldsymbol{\rho}_R,\Phi_L,\Phi_R\}$.
The maximization over these is a straightforward approach closely linked to the single detector approach. Towards the end, we connect
all the previous works related to the multi-detector Gravitational Wave inspiral search and express in the same notation in order to bring under the same footing. 
\end{abstract}


\maketitle

\section{Introduction}

Global network of broad band advanced gravitational wave(GW) detectors such as Advanced LIGO, Advanced Virgo \cite{ALIGO,AVIRGO} will be ready
in next few years. 
Japanese detector KAGRA is under construction and would be functioning in next decade \cite{KAGRA}. The
detectors are aimed to detect few inspiral events per month from one of the
prominent sources of GWs namely compact binaries with a range around
200 Mpc when GW search is carried out individually on each detector \cite{EventRate,ALIGO}.

The compact binary (with component masses $m_1$ and $m_2$) search through the detection of inspiral phase is carried out by phase matching technique well known as the matched filtering applied to the
output data from a single detector. The signal is characterized by 
many physical parameters namely, for non-spinning inspiraling binaries the parameter space is $\ll({\cal M}, \eta, \text{A}_0 ,t_a,\phi_a \rr)$. Here, $\phi_a$
is the phase of the signal at the time of arrival $t_a$. The mass ${\cal M}=(m_1m_2)^{3/5}/(m1+m2)^{1/5}$ is the chirp mass and $\eta = m_1m_2/(m_1+m_2)$ is the  symmetric mass ratio.
$\text {A}_0$ is the overall amplitude. The detection
is carried out by laying the templates in the multi-dimensional parameter space and then
maximizing the matched filter output. This is technically known as Maximum Likelihood Ratio (MLR) analysis which is assumed to be optimal when signal-to-ratio
is large and signal is buried in the noisy Gaussian data \cite{Owen:1998dk,sanjeev1991}.

In inspiral search using a multi-detector network with Advanced LIGO, Advanced Virgo etc, each interferometer is arbitrarily oriented due to its location on the globe. Thus they have different response to the incoming GW from given direction. Thus, the phase coherent detection
with multi-detector network which would involve incorporating the inspiral signal from
different detectors in a phase coherent fashion is a natural way to carry out multi-detector coherent analysis. Further, with this additional information from
different detectors, the network removes the degeneracy of the single detector.
As a result the search parameter space increases to $\ll({\cal M}, \eta, \text{A}_0,t_a,\phi_a,\theta,\phi,\epsilon,\Psi \rr)$. Where $(\theta,\phi)$ represents the source location,
 $\epsilon$ is the inclination angle  and $\Psi$ is the polarization angle of the binary system. For example, two independent detectors can
measure GW
polarization predicted by Einstein's GR and hence the polarization becomes an explicit parameter. Similarly, at least 4 detectors are necessary to
measure the source location by triangulation method\cite{Triangulation}. 

The multi-detector inspiral GW search has been developed in the literature
for more than a decade by various groups around the world \cite{apaiprd01, finn2000, harryprd11}. 
In \cite{apaiprd01}, the coherent formalism based on the MLR was developed and the 4 out of above mentioned 9 parameters, namely $(\text{A}_0,\phi_a, \epsilon,\Psi)$ 
were maximized using the rotation group symmetries and Gel-Fand functions. The network likelihood ratio was shown to be sum-square of
the projected network correlation function vectors on the (2-D) polarization plane. In \cite{finn2000}, the coherent formalism
was developed for a general case of correlated noise (to treat LIGO-Hanford and LIGO-Livingston). In \cite{harryprd11}, the coherent formalism was developed based on the F-statistic \cite{Jaranowski1998,PulsarNetwork}, where the 4 physical
parameters ${\ll(\text{A}_0,\phi_a, \epsilon,\Psi \rr)}$ are mapped to the 4 amplitude parameters namely $(\cA_1, \cA_2, \cA_3, \cA_4)$. 
It was shown that the MLR over these parameters is equivalent to the linear least square problem. The authors showed 
that the freedom of polarization angle $\Psi$ allows to set the dominant polarization frame in which the two GW polarizations are separable. 
For more reference on the dominant polarization frame; see \cite{klimenkoprd05}. Finally, the authors show that the MLR
analysis gives two terms which can be interpreted as the matched filtering output of the 
two synthetic streams constructed from the output of the multiple detectors; e.g. Eq.(2.35) of \cite{harryprd11}. For more references on the
synthetic streams in the GW burst context; see \cite{sylvestre2005} and GW unmodeled chirps; see \cite{apaiprd06}.

Synthetic streams would act like the building blocks for the Aperture Synthesis technique for GW search with two polarizations predicted by Einstein's GR. 
Aperture Synthesis is a standardized technique to combine signals from multiple telescopes for localization in electromagnetic window. This technique is used in the optical as well as in the radio telescopes
where the signals are combined from distinct telescopes with the directional dependent coefficients to point to a direction. The earlier literature in inspiral
search using multiple detectors  \cite{apaiprd01, finn2000, harryprd11} shows that the MLR in the multi-detector framework gives two synthetic data streams.
The quadrature matched filtering through these synthetic data streams gives the network likelihood.

In this paper, we show that the synthetic streams is a natural way to combine the data streams in a phase coherent fashion mimicking  a pair of 
{\it effective multi-detector antennas} even before the construction of MLR. This is obtained using the multi-detector 
singular-value-decomposition (SVD) of the {\it 'signal-to-noise ratio (SNR) vector'} in a much more straightforward way. The highlights of the paper are as follows. 
\begin{enumerate}
\item
The singular vectors of the {\it network SNR vector} naturally gives the {\it Dominant Polarization (DP)} frame.
\item
The new formalism imposes the simultaneous matched filtering in the spectral as well as network space and
hence termed as {\it network matched filter} which give a pair of synthetic streams, which is similar to the ones obtained in earlier literatures through Maximum Likelihood Ratio(MLR). 
\item
The network log likelihood ratio (LLR) expressed in terms of the synthetic streams corresponding to the two singular vectors is the sum
of the LLRs of the two effective synthetic data streams. 
\item
In this framework, the 4 physical extrinsic parameters, $(\text{A}_0,\phi_a, \epsilon,\Psi)$  are mapped to four effective parameters, namely
$\ll(\boldsymbol{\rho}_L,\boldsymbol{\rho}_R, \Phi_L,  \Phi_R \rr)$.
\item
The network MLR amounts to maximizing LLR over the 2 effective amplitudes $\ll(\boldsymbol{\rho}_L,\boldsymbol{\rho}_R\rr)$ and 2 effective phases,
$\ll(\Phi_L,  \Phi_R \rr)$. This is a clear extension of the single detector data stream to analysis of two effective synthetic streams (due to multi-detector network)
 corresponding to two polarizations. Thus, the synthetic streams mimic the multi-detector network. 
\end{enumerate}

The approach gives an elegant way to obtain and construct the synthetic streams in multi-detector analysis. Further, we connect this work to all the existing works namely
\cite{apaiprd01, harryprd11} bringing different notations under the same umbrella.


The paper is organized as follows. Sec. \ref{sec:BinarySignal} gives a brief outlines the binary signal 
with two polarizations in the multi-detector network. In Sec. \ref{sec:NetSNRSec} we define a SNR vector 
$\boldsymbol{\varrho}$ and using the singular-value decomposition technique, we show that the norm of the same which is the network
SNR is the square root of the sum of the two distinct  
quantities, $\boldsymbol{\rho}_L^2$ and $\boldsymbol{\rho}_R^2$. In Sec. \ref{sec:synthstreams} we obtain two synthetic streams as linear 
combinations of network signal such that their respective SNR's are $\boldsymbol{\rho}_L$ and $\boldsymbol{\rho}_R$. In Sec. \ref{sec:LR},
we construct the network Likelihood statistic, $\varLambda$ in terms of synthetic streams. Further, we show that the extrinsic physical
parameters are mapped into $\ll(\boldsymbol{\rho}_L,\boldsymbol{\rho}_R, \Phi_L,  \Phi_R \rr)$. {\it Thus, the network MLR analysis
is equivalent to the MLR analysis of the 2 independent network synthetic streams and the maximization is equal to maximizing the 2 effective
amplitudes and the 2 effective phases.} In Sec. \ref{literaturecomp} we connect our results to the existing literature. 

\section{GW Inspiral Binary Signal in a Multi-Detector Network}
\label{sec:BinarySignal}
In this section, we construct the inspiral GW signal as observed by a network
of $I$ interferometric GW antennas. The time domain restricted GW waveform (two polarizations) from
a non-spinning compact binary with masses $m_1,m_2$ and located at a distance $r$ 
which enters the interferometric detector band at time $t_a$ with the phase $\phi_a$ is
\bea
 h_+(t) &=& \text{A}(m_1,m_2,r,t) \frac{1+\cos^2 \epsilon}{2}  \ \cos \ll[\Phi(t-t_a)+\phi_a \rr]\, , \nonumber \\
 h_\times (t) &=& \text{A}(m_1,m_2,r,t) \cos \epsilon \ \sin \ll[\Phi(t-t_a)+\phi_a \rr] \,, \label{eq:hp}
\eea
where $\Phi(t-t_a)$ is the phase restricted to 3.5 PN order and  the inclination angle, $\epsilon$ is the
angle between orbital angular momentum vector and observer's line of sight.

 The strain (response) measured by any interferometric GW detector to this GW is
\bea
 s(t) \equiv  {\text{F}_+}~h_+(t) + {\text{F}_\times}~h_\times(t) \, , \label{eq:tsignal1}
\eea
where ${\text{F}_+}$ and ${\text{F}_\times}$ are the antenna pattern functions which describe the
angular response of an antenna to a given source location. Thus, they depend on the source location with respect
to the detector's site.

In a multi-detector network with $I$ antennas at distinct locations, we represent the antenna patterns as $I$-dimensional vectors
namely ${\bf F_{+,\times}} =\{ {\text{F}_{+,\times}}_m \}$  where the subscript $m$ varies from 1 to $I$. In addition,
different detectors receive the ${h_{+,\times}}_m$ with time-delays depending on the location of the detector on Earth's globe.
Thus, the signal at the $m$-th detector site is $s_m(t)=s_{ref}(t-\tau_m)$, where $s_{ref}(t)$ is the GW signal in the geocentric
frame, which we treat as reference here and $\tau_m$ is the time delay between signal in the  $m$-th detector and the geocentric frame with respect to source location. However, for the context of this
paper, henceforth we assume that we compensate for the delays and consider the delayed signal keeping the same notation to
construct the network signal\footnote{Note that for construction of the likelihood, this is reasonable. However, if one needs
to place the templates in the directional space\cite{SkyTemplateKeppel}, one needs to explicitly incorporate the delays in the formalism.}.

The antenna pattern vectors, ${\bf F_{+,\times}}$ are the functions of the polarization angle $\Psi$, source location $(\theta,\phi)$ and detector's Euler angles  with respect to geocentric frame.
We use $(\alpha_m,\beta_m,\gamma_m)$ to represent $m$-th detector's Euler angles following the convention in the literatures \cite{apaiprd01,apaiprd06,harryprd11}.

In the same spirit, we express the delayed and sampled signal arrived at the $m$-th detector with $2N$ number of time
samples as a $2N$ dimensional vector,\footnote{We take $2N$ time samples so that there would be $N$ positive frequency samples as we work in
frequency domain rest of the paper} 
\be
{\s_m} = {\text{F}_+}_m~\h_+ + {\text{F}_\times}_m~\h_\times \equiv \Re \ll[ \text{F}_m^*~ \h \rr] \, , \label{eq:tsignal2}
\ee
where $\h= \h_+ +i \h_\times$ and  $\text{F}_m ={\text{F}_+}_m + i {\text{F}_\times}_m $ is the complex antenna pattern.

In frequency domain signal becomes,
\be
{{\mathbf{\tilde s}}_m} = {\text{F}_+}_m~\th_+ + {\text{F}_\times}_m~\th_\times \, , \label{eq:fsignal}
\ee
where $N$-dimensional vectors $\th_+$ and $\th_\times$ are the discrete versions of frequency domain GW polarizations
$\tilde h_+(f)$ and $\tilde h_\times (f)$ in the positive frequency region. From  Eq.\eqref{eq:hp}, it can be shown\footnote{The Fourier transform is, $\tilde G(f) = \int_{-\infty}^\infty G(t) e^{-2 \pi i ft}dt.$}, that
 \bea
\tilde h_+(f)&=&\text{A}_0 \frac{1+\cos^2 \epsilon}{2} ~h_0(f) e^{i \phi_a}~,  \nonumber \\
\tilde h_\times(f)&=& \text{A}_0 \cos \epsilon ~h_{\pi/2}(f) e^{i \phi_a}~,
\eea
with $\tilde h_0(f)=i \tilde h_{\pi/2}(f)= f^{-7/6} e^{i \varphi(f)} $ , the stationary phase approximated frequency domain waveform. 
Here $\text{A}_0$ is the constant which depends on the masses and the distance and $\varphi(f)$ is the
 $3.5$ PN corrected phase of inspiral signal\cite{arun04}. In the coming sections we use $\tilde \h_0$ and $\tilde \h_{\pi/2}$ to represent discrete versions of $\tilde h_0(f)$ and $\tilde h_{\pi/2}(f)$
respectively.

Thus, the incoming signal to a multi-detector network of $I$ interferometric detectors with $N$  number of
frequency samples is expressed as a $N\times I$ matrix,
\bea
 \boldsymbol{\tilde \S}_{N \times I} &\equiv& [\mb{\tilde s}_1~ \mb{\tilde s}_2 ~\ldots \mb{\tilde s}_I]\,.  \label{eq:fsignalnet}
\eea

\section { Network SNR}
\label{sec:NetSNRSec}
We assume that the noises in individual detectors to be  independent, additive, stationary Gaussian. Thus the
square of matched filter SNR of the combined network is the sum of squares of matched filter SNRs of the
individual detectors \cite{apaiprd01,apaiprd06,harryprd11} as follows,
{\small \bea
\boldsymbol{\rho}^2 &=& \sum_{m=1}^I \rho_m^2 =  4 \sum_{m=1}^I   \ll[ \sum^N_{j=1} \frac{\ll|\tilde \S_{jm} \rr|^2}{\N_{jm}} \rr] \, , \nonumber \\
&=& \text{A}^2_0 ~~\ll[\sum_{m=1}^I g_m^2 {\text{F}^2_+}_m \ll(\frac{1+\cos^2 \epsilon}{2} \rr)^2 +   g_m^2{\text{F}^2_\times}_m \cos^2\epsilon\rr],\label{eq:snrnet1}
\eea}
where, $\N_{jm}$ is the $j$-th frequency component
 of  one sided noise power spectral density(PSD) vector of $m$-th antenna\footnote{$ {\bf E} \ll[| \tilde n_{jm} |^2 \rr]=\frac{1}{2}\N_{jm}$, where  $\tilde n_{jm}$
 is the $j^{th}$ frequency component of noise in the $m^{th}$ detector.}\cite{Schutz:1997bx} and  $g_m^2 = \langle \h_0|  \h_0 \rangle_m$
\footnote{The scalar product $\langle \mathbf{a} |\mathbf{b} \rangle_m =  4 \Re \sum_{j=1}^N \frac{ {\tilde a}_j~ {\tilde b}_j^*  }{\N_{jm}} \equiv 4 \Re \sum_{j=1}^N \tilde{\tilde a}_j~ {\tilde b}_j^*$. \\
$\mathbf{\tilde{\tilde a}}$ is the frequency domain vector for the over-whitened signal $\mathbf{a}$.}.
Here  subscript $m$ denotes the noise PSD of $m$-th detector to be used. The $g_m$'s depict the SNR ratio in different detectors
arising solely due to the noise PSD. When all the noise PSD's are identical then $g_m=g$, a constant. Note that this $g_m$ is proportional to the one
defined in \cite{apaiprd01}-Eq.(3.12).
\subsection{Network SNR Vector}

The form of $\boldsymbol{\rho}^2$ in Eq.\eqref{eq:snrnet1} motivates us to define a $I$-dimensional SNR vector,
\bea 
\boldsymbol{\varrho} \equiv \text{A}_0 \ll[ \ll(\frac{1+\cos \epsilon}{2}\rr)^2 e^{-2i\Psi}~~~\ll(\frac{1-\cos \epsilon}{2}\rr)^2 e^{2i \Psi} \rr] 
\underbrace{\begin{bmatrix} {\d}^T \\ \\ {\d}^H \end{bmatrix}}_{\mb{D}^T} , \label{eq:rho1}
\eea
with $\d \equiv \{ \F' \exp(2 i \Psi) \}$ and $\d^T$, $\d^H$ represent Transpose and  Hermitian conjugate of $\d$ respectively, where
\be \text{F}'_m=g_m \text{F}_m= d_m(\theta,\phi,\alpha_m,\beta_m,\gamma_m) e^{-2 i \Psi} \, ,
\label{eq:apattern}
\ee
is the noise weighted complex antenna pattern function constructed from ${\text{F}_+}_m$, ${\text{F}_\times}_m$ and $g_m$ \footnote{The vector 
$\d$ is the noise weighted version of the one defined in \cite{apaiprd06}}. Further, it can be easily shown that the norm-square
$\boldsymbol{\varrho} \boldsymbol{\varrho}^H$ is the network SNR square $\boldsymbol{\rho}^2$ as given in Eq.\eqref{eq:snrnet1}.

Here we use the fact that the $\Psi$ dependence in the complex antenna pattern can be separated into a phase
as given in Eq.\eqref{eq:apattern}. This ensures freedom in the choice of $\Psi$ {\it via} the orientation of the wave frame with respect
the reference frame. If we rotate the wave plane by an additional angle $\Delta \Psi$ about the line of sight(Z-axis), the network antenna
pattern for this newly rotated wave frame can be obtained by applying an equivalent transformation on {\F} as, $\F \exp{(-2 i \Delta \Psi)}$. 
That would also transform the signal as $\h \exp{(-2 i \Delta \Psi)}$ keeping the response Eq.\eqref{eq:fsignal} invariant.
This freedom in the choice of $\Psi$ is crucial to choose appropriate frame.

We use the technique of SVD to decompose $\boldsymbol{\varrho}$ into a sum of two orthogonal vectors in the $I$-dimensional
space. We further show below that the SVD naturally sets a particular choice for $\Psi$ which is indeed shown to be related to the DP frame discussed in the literature.

The SVD of $\D$ is,
\bea
 \mb{D}
 = \underbrace{[{\hu_1} \quad {\hu_2}]}_{\U} 
 \ \underbrace{\begin{bmatrix}  \frac{\| \u_1 \|}{\sqrt{2}} &  &0  \\ 0&   & \frac{\| \u_2 \|}{\sqrt{2}}  \\  \end{bmatrix}}_{\Sigma}
 \underbrace{\frac{1}{\sqrt{2}} \begin{bmatrix} e^{i \frac{\delta}{2}}&  & e^{-i \frac{\delta}{2}} \\i e^{i \frac{\delta}{2}}&  &-i e^{-i \frac{\delta}{2}} \end{bmatrix}}_{\V^H} \, ,
\eea
where $\ll(\U, \Sigma,\V \rr)$
 have similar form as described in section \Rmnum{4} of \cite{apaiprd06}. 
The columns of $\U$, $\hat \u_1$ and $\hat \u_2$ are the left singular vectors and  those of $\V$  are  the right singular
vectors of $\D$. The diagonal elements of $\Sigma$ are  the singular values of $\D$. 

The orthogonal pair $\{\u_1, \u_2\}$ can be written down in terms of the antenna pattern functions as
\bea
\u_1 &\equiv& 2 \Re \ll[ \F' e^{2i \chi}\rr], ~~~~~~~ \u_2 \equiv 2 \Im \ll[ \F' e^{2i \chi}\rr] \, , \label{eq:u1u2} 
\eea
with\footnote{$\|\ . ~\|$ represents the Euclidean norm of a vector.} $\|\u_{1,2}\| = \sqrt{2\ll(\d^H \d \pm |\d^T \d |\rr)}$ and
the phase $\chi = \Psi-\frac{\delta}{4}$. The shift in $\Psi$,  $\delta/4 = arg(\d^T \d)/4$  solely depends on the multi-detector
network orientation with respect to the source location. Thus, for different sky-positions, the $\chi$ would vary for a given network
configuration.

Substituting back in Eq.\eqref{eq:rho1}, we obtain the network SNR vector as a linear combination of two orthogonal vectors
{\small \bea
\boldsymbol{\varrho} &=& \frac{\text{A}_0}{2}\ll( \frac{1+\cos^2 \epsilon}{2} ~ \cos 2\chi  - i \cos \epsilon~ \sin2 \chi \rr) \u_1^T \nonumber \\
&+& \frac{\text{A}_0}{2} \ll( \frac{1+\cos^2 \epsilon}{2}~  \sin 2\chi  + i  \cos \epsilon ~\cos 2 \chi \rr) \u_2^T \, . \label{eq:rho2}
\eea}

By using orthogonality property of $\u_1$ and $\u_2$, we can show that  $\boldsymbol{\rho}^2$  can be written as the sum of two 
individual terms arising from the orthogonal vectors in the network SNR vector. Please note that unlike Eq.\eqref{eq:snrnet1},
 the above equation is expressed in terms of orthogonal $\{\u_1,\u_2\}$ pair.
For the sake of completeness, we give below the network SNR in terms of the individual SNRs.
{\footnotesize \bea
\boldsymbol{\varrho} \boldsymbol{\varrho}^H &=& \boldsymbol{\rho}_L^2 +\boldsymbol{\rho}_R^2 \nonumber \\
&=&  \frac{\text{A}^2_0 \|\u_1\|^2}{4} \underbrace{\ll[ \ll(\frac{1+\cos^2 \epsilon}{2} \rr)^2 \cos^2 2 \chi + \cos^2 \epsilon \sin^2 2 \chi \rr]}_{\text{P}^2_L} \nonumber \\
&+&  \frac{\text{A}^2_0 \|\u_2\|^2}{4} \underbrace{ \ll[ \ll(\frac{1+\cos^2 \epsilon}{2} \rr)^2 \sin^2 2 \chi + \cos^2 \epsilon \cos^2 2 \chi \rr]}_{\text{P}^2_R} \label{eq:snrnet2}
\eea}  

Here we can see that the two linear polarizations, $(+,\times)$ are linearly combined to form a pair of Left(L) and Right(R) circular polarizations. More discussion on circular polarizations is given in section \ref{sec:synthstreams}.

This motivate us to construct two synthetic streams which would each give an individual SNR,  $\boldsymbol{\rho}_L$ and $\boldsymbol{\rho}_R$. We address this in the subsequent section \ref{sec:synthstreams}.


\subsection{Connection to Dominant Polarization Frame}
Dominant polarization(DP) frame is a specific choice of wave frame in which the the real and imaginary
part of noise weighted complex network antenna pattern vector, $\text{\bf{F}}'^{DP}=\text{\bf{F}}'^{DP}_+ + i\text{\bf{F}}'^{DP}_\times $ 
becomes orthogonal to each other. {\it i.e.}
\bea
\sum_{m=1}^I {\text{F}'_+}_m^{DP} {\text{F}'_\times}_m^{DP} =0. \label{eq:dpant0}
\eea
This is possible by a certain choice of polarization angle, $\Psi^{DP}$ of wave frame with respect to geocentric
frame . The word was coined in the context of detection of bursts by Klimenko \cite{klimenkoprd05}  and recently in
the inspiral search with multi-detector network in \cite{harryprd11}.

From Eq.\eqref{eq:apattern}, we can write 
\bea
\F'^{DP} = \d ~ e^{-2 i \Psi^{DP}}.
\eea
Our aim is to relate $\Psi^{DP}$ to the angles given in the SVD framework.


We note that a vector $\F' e^{2i \chi}$ from Eq.\eqref{eq:u1u2} has orthogonal real and imaginary
component vectors, which is precisely the condition for antenna pattern vectors in the
DP frame. Thus, the SVD provides a natural connection to the DP frame through its construction.

In summary, 
\bea \F'^{DP}&=& \text{\bf F}' e^{2i \chi } =\frac{\u_1+i \u_2}{2}= \d e^{-i \frac{\delta}{2}}~, \nonumber \\
\F'^{DP}_+ &=&  \frac{\u_1}{2} ~~~~\text{and}~~~\F'^{DP}_\times = \frac{\u_2}{2} \label{eq:fnew}~ \label{eq:dpant1} ,\eea
and 
$\Psi^{DP}=\delta/4$. The DP frame is obtained by rotating wave frame about z-axis by an angle ~$\chi = \Psi - \delta/4 $ in
clockwise direction. As is expected, the DP frame is pertaining to a source; {\it i. e.} with the change in the location of
the source, the $\chi$ (the angle through which one needs to rotate to bring into the DP frame) would change.

\section{Aperture Synthesis in GW inspiral search}
\label{sec:synthstreams}
As mentioned earlier, more than one detector is necessary to determine GW polarization as well
as localization of the sources \cite{tintoprd89}.
In the following discussion, we construct two effective synthetic data streams out of $I$ 
detector data streams in the network, they together carry full network $SNR$ as given in Eq.\eqref{eq:snrnet1}.
Since GW carry two polarizations in the Einsteinian GR, the signal resides in the 2-dimensional sub-space of $I$-dimensional
network space and hence only two independent data streams are sufficient to provide information about the polarization
in the network formalism. We show that the constructed data streams provide that information. In the past 
\cite{apaiprd01,apaiprd06,harryprd11}, the synthetic data streams were shown to be the by-product of maximizing the network LLR over
the four extrinsic parameters $(A, \phi_a, \epsilon, \Psi)$. However, here in this section, we show that the synthetic streams can be constructed prior to
the MLR analysis. This is similar to the spirit of aperture synthesis technique in the electromagnetic window such as optical or radio used
in Very Large Telescope Interferometer(VLTI) \nocite{VLTI} or Very Long Baseline Interferometry (VLBI) \nocite{VLBI}, where an effective antenna is constructed out of linear combination of data from different telescopes.

First, we construct the synthetic data streams using $\u_1$ and $\u_2$ and show that
they individually give the matched filter SNR equal to $\boldsymbol{\rho}_L$ and $\boldsymbol{\rho}_R$.

\subsection{Signal $\s_m$ in the DP Frame}
In the rest of the paper, we work in the new frame provided by the SVD {\it aka} the DP frame.
We first rewrite the antenna pattern {\bf F} in terms of
$\u_1$ and $\u_2$, and then express the network signal in time as well as
frequency domain as below.

Using Eq.\eqref{eq:tsignal2} and Eq.\eqref{eq:fnew}, the time domain signal vector at $m$-th detector is 
{\small \bea
\s_m  &=& \frac{1}{2} ~\Re \ll[ \h~ e^{2i \chi} ~\frac{{u_1}_m}{g_m} \rr]   + \frac{1}{2} ~\Im \ll[ \h e^{2i\chi} ~\frac{{u_2}_m}{g_m} \rr]  , \nonumber\\
&=& \Re \ll[ \h~ e^{2i \chi} ~\ll(\frac{{u_1}_m+i {u_2}_m}{2g_m}\rr)^* \rr] \equiv \Re [\h^{DP}~ \F^{DP*}_m] \, .
\eea}
where $\h^{DP} = \h \exp(2 i \chi)$ and $\F^{DP} = \F \exp(2 i \chi)$ are the complex GW  as well as the network antenna pattern function in the DP frame respectively.

In frequency domain, the $j$-th frequency component of the signal in $m$-th detector, $\ll[ \tilde \s_m\rr]_j = \tilde \S_{jm}$
can be expressed in terms of linear combination of {\small $\F^{DP}_{+,\times}$} in the DP frame. Further, the amplitude $A_0$,
initial phase $\phi_a$ and frequency dependent part namely $\tilde{\h}_0$ are factored out as shown below
{\small \bea
\tilde \S_{jm}  &=&  \text{A}_0 e^{i \phi_a} \tilde{h}_{0j}  \ll [\ll(\frac{1+\cos^2 \epsilon}{2}~ \cos 2 \chi + i \cos \epsilon~ \sin 2 \chi \rr) \text{F}_{+ m}^{DP} \rr. \nonumber \\
&&+ \ll.\ll(\frac{1+\cos^2 \epsilon}{2}~ \sin 2 \chi -i  \cos \epsilon~ \cos 2 \chi \rr) \text{F}_{\times m}^{DP} \rr]  \nonumber \\
&\equiv& \text{A}_0 \tilde{h}_{0j}  \ll [ \text{P}_L \, \text{F}_{+ m}^{DP}\, e^{i \Phi_L}  + \text{P}_R \,\text{F}_{\times m}^{DP}\, e^{i \Phi_R} \rr] \, . \label{eq:sij}
\eea}
The $\text{P}_{L,R}$ are the polarization amplitudes in the DP frame as defined in Eq.\eqref{eq:snrnet2}.
This carry the effect of rotation by $2 \chi$ of the signal which mixes the two linear GW polarizations into a pair of  {\it left(L) and right(R) Circular Polarizations}.
The polarization phases are,
\bea
\Phi_L(\epsilon,\chi) &=& {\rm tan^{-1}}\ll[{\rm tan}(2 \chi) ~~ \frac{2\cos \epsilon}{1 + \cos^2 \epsilon} \rr] + \phi_a, \\
~~\Phi_R(\epsilon,\chi) &=& \Phi_L \ll(\epsilon,\chi+\frac{\pi}{4}\rr).
\label{eq:polph}
\eea
As expected the $\Phi_R$ phase is obtained by rotating $\chi$ by $\pi/4$ in $\Phi_L$; the
property of GW polarizations. The above polarization terms; namely 
$\{\text{P}_{L,R} e^{i \Phi_{L,R}}\}$ can be shown to be equal to 
\bea
\text{P}_{L} e^{i \Phi_{L}}\ &=& \ll[T^*_{2+2}(\chi,\epsilon,0)+ T^*_{2-2}(\chi,\epsilon,0)\rr] \\
\text{P}_{R} e^{i \Phi_{R}}\ &=& i \ll[T^*_{2-2}(\chi,\epsilon,0)- T^*_{2+2}(\chi,\epsilon,0)\rr]
\eea
respectively, which describe the circular polarizations expressed in terms of  rank-2 Gel-Fand function $T_{mn}$ \cite{Dhurandhar88} \footnote{The rank-2 Gel-Fand functions
$$T^2_{2\pm2}(\chi,\epsilon,0) = \frac{\ll(1 \pm \cos \epsilon \rr)^2}{4} \exp{(\mp 2 i \chi)}.$$ Since throughout the paper we use only rank-2 Gel-Fand functions, we drop the superscript 2 from $T^2_{mn}$.}.

In the next subsection, we use this separation feature of the signal to construct the synthetic streams.

\subsection{Network Matched Filter}
In this section, we introduce the notion of a matched filter designed for a the multi-detector analysis which not only combines the spectral but also the network features. We call this filter as the {\it network matched filter}. 

Let the frequency domain delayed network data stream is given by
$ \boldsymbol{\tilde \X}= [\tilde \x_1~\tilde \x_2.~.~.~\tilde \x_I]$ with $\tilde \x_m= \tilde \s_m +\tilde \n_m$ and $\tilde \n_m$ is the frequency domain
noise vector corresponding to $m$-th detector. To proceed further, we make following constructs.
\begin{enumerate}
\item
{\it Over-whitened data stream}: Construct over-whitened data stream 
incorporating the noise PSD of the individual antennas 
denoted by ${\tilde {\tilde \X}}_{jm} = \frac{\tilde \X_{jm}}{\N_{jm}}$.
\item
{\it Synthetic data stream}: The over-whitened synthetic data stream $\tilde {\tilde \z}$ is constructed from the linear combination of individual over-whitened data streams as below,   
\be
\tilde {\tilde z}_j \equiv \sum_{m=1}^I \alpha_m \tilde {\tilde \X}_{jm},\label{eq:z}
\ee
with real linear coefficients $\alpha_m$. The over-whitened data is used for the synthetic data stream construction in order to
incorporate the individual noise PSDs.  
\end{enumerate}
We show in the rest of the section that using the classical idea of matched filtering, we can tune
$\alpha_m$ such that the resulting synthetic data stream would observe either $left$(L) or $right$(R) circular polarization.

In the next subsection, we remind the reader the classical derivation of the
matched filter used for the single detector context.

\subsubsection{Single Detector Matched Filter}
If $\tilde \k$ is a filter, then the filtered output of $\tilde \z$ through this filter is $ \langle  \z~ |~  \k \rangle$. Here,
(in order to avoid the excess notations) for this sub-subsection, let us assume that $\tilde \z$ denotes
the un-whitened data of the single detector. 

The SNR of the filtered output is \cite{maggiore},
{\footnotesize \bea
{\rm SNR} &=&\frac{\text{\bf E} \ll[\langle  \z~ |~ \k \rangle \rr]}{{\boldsymbol{\sigma}} \ll. \ll[ \langle  \z~ |~ \k \rangle \rr] \rr|_{\boldsymbol{\tilde \S}=0}} 
= \frac{ 4 \Re \ll[ \sum_{j=1}^N \text{\bf E} \ll(\tilde{\tilde{z}}_j \tilde {k^*}_j \rr) \rr]}{ \sqrt{\text{E} \ll. \ll[  \ll( 4 \Re \ll[ \sum_{l=1}^N \tilde{\tilde{z}}_l {\tilde k^*}_l \rr] \rr)^2 \rr] \rr|_{\boldsymbol{\tilde \S}=0}}},\nonumber \\
 \label{eq:MfilterSNR1}
\eea} 
where $\text{\bf E[~.~]}$ represents the expectation and $ {\boldsymbol{\sigma}}\text{\bf [~.~]}$ represents the standard deviation.
The single detector SNR further simplifies to 
\be \text{Single detector SNR} = \frac{4 \Re \ll[  \sum_{j=1}^N
 \tilde{\tilde s}_{j}~  \tilde k^*_{j}  \rr]}{\sqrt{4 ~\sum_{j=1}^N \frac{\ll|\tilde{k}_{j}\rr|^2}{\N_{j}}}}
= \langle \s | \hat{\k} \rangle .
\label{eq:singleSNR}
\ee
The filter norm is ${\sqrt{\langle \k| \k \rangle}}$. As is known, the above SNR would be optimal when the filter vector is aligned
to the signal vector {\it i.e.} $\tilde k_j \propto \tilde s_j$ known as the matched filter.

Now, let us explore these ideas in the context of the multi-detector scenario.

\subsubsection{Network Matched Filter}

The matched filter is that filter which gives the optimum SNR in Gaussian noise.  However, the signal in DP frame
is separated in such a way that the noise weighted antenna patterns are orthogonal. Let us apply the matched filter notion with the aim that
the resultant combined spectral as well as network filter {\it via} $\alpha_m$ would capture the individual polarizations. We show below that
in this exercise, this amounts to constructing a combined matched filter.

Consider Eq.\eqref{eq:MfilterSNR1} with $\tilde \z$ as the synthetic stream defined 
in Eq.\eqref{eq:z}. Then, Eq.\eqref{eq:MfilterSNR1} can be simplified to  
\bea
{\rm SNR} = \frac{4 \Re \ll[  \sum_{j=1}^N \sum_{m=1}^I  \tilde{\tilde \S}_{jm} {\tilde \K^*}_{jm} \rr]}
{\sqrt{4 ~\sum_{j=1}^N \sum_{m=1}^I   \frac{\ll|\tilde{\K}_{jm}\rr|^2}{\N_{jm}}}}~, \label{eq:MfilterSNR2}
\eea
where $\boldsymbol{\tilde \K} = \tilde \k \otimes \boldsymbol{\alpha}$. {\it i.e.} $\tilde \K_{jm}= \tilde k_j \alpha_m$ and
$\tilde \K_m = \alpha_m \tilde \k$.

Further The denominator of Eq.\eqref{eq:MfilterSNR2},
\be
\sqrt{4 ~\sum_{j=1}^N \sum_{m=1}^I   \frac{\ll|\tilde{\K}_{jm}\rr|^2}{\N_{jm}}} = \sum_{m=1}^I \alpha_m^2 \langle  \k|  \k \rangle^2_m \label{eq:filternorm}
\ee
is the Frobeinus matrix norm of  $\bf{\tilde{\K}}_{N \times I}$ in the combined spectral-network ($N \times I$) space
defined as $||{\tilde{\K}}||^2 \equiv {\rm Tr}(\langle \tilde{\K_m}, \tilde{\K_n} \rangle)$ which is same as the right hand side of
Eq.\eqref{eq:filternorm}. The subscript $m$ in $ \langle  \k|  \k \rangle_m$ denotes the noise PSD is that of $m$-th detector.

It is interesting to note that, Eq.\eqref{eq:MfilterSNR2} has the same structure of a conventional single detector 
matched filter SNR in the Eq.\eqref{eq:singleSNR}. For a single detector, filter is a one dimensional vector in the spectral direction. As the matched filter is that filter which gives maximum SNR, which should be aligned along the signal {\it i.e.}
$ {\tilde k}_j  \propto {\tilde s}_j$ for single detector case. 

While in the multi-detector case, $\boldsymbol{\tilde{\K}}$ is a 2-dimensional $(N \times I)$ combined  spectral-network filter.
Since $\boldsymbol{\tilde \S}$ can be decomposed into frequency and network components, the optimal combined filter
should {\it match} with $\boldsymbol{\tilde \S}$ (or part of $\boldsymbol{\tilde \S}$) in the same spirit as that
of the single detector case described above. Owing to the fact that  detailed in Eq.\eqref{eq:sij}, we recall that
the network signal has two constituent parts. Here, we construct that filter which captures either of the two circular polarizations as shown below.

\begin{enumerate}
 \item {\bf $\boldsymbol{\tilde \K}_L$: Network Matched Filter for Left Circular polarization} \\
To capture the plus polarization of DP frame ($Left$ circular polarization) in Eq.\eqref{eq:sij}, $\boldsymbol{\tilde{\K}}_L$ should be
aligned to the $Left$ polarization part of the $\boldsymbol{\tilde \S}$ which we call
$\boldsymbol{\tilde \K}_L \equiv \tilde \k_{L} \otimes \boldsymbol{\alpha}_{L}$.
The alignment condition demands that it should satisfy     
\be
{\tilde k}_{L ,j} \propto {\tilde h}_{0j} e^{i \Phi_L}~~~\text{and}~~ {\alpha_{L,m}} \propto \text{F}_{+m}^{DP} \, .\label{eq:filter11}
\ee
together. Since the Frobeinus {\it norm} of $\boldsymbol{\tilde \K}_L$ is $\| \F'^{DP}_+\|$, the components of normalized network plus filter $\boldsymbol{\tilde \K}_L$ become,
{\be
\tilde k_{L,j} = \tilde h_{0 j} e^{i \Phi_L}~, \hspace{0.2in} \alpha_{L,m} =  \frac{\text{F}^{DP}_{+m}} { \| \F'^{DP}_+ \|}~, \label{eq:filter12}
\ee}
Using Eq.\eqref{eq:MfilterSNR2} and Eq.\eqref{eq:filter12}, the corresponding SNR becomes, 
{\be
4 \Re \ll[  \sum_{j=1}^N \sum_{m=1}^I \tilde {\tilde \S}_{jm}  ~\tilde \K_{L,jm}  \rr] = \boldsymbol{\rho}_L~. \label{eq:snr1}
\ee}

\item {\bf $\boldsymbol{\tilde \K}_{R}$: Network Matched Filter for Right Circular polarization}:~ \\
To capture the cross polarization of DP frame ($Right$  circular polarization) in Eq.\eqref{eq:sij}, we construct another filter,
$\boldsymbol{\tilde \K}_R \equiv \tilde \k_{R} \otimes \boldsymbol{\alpha}_{R}$  such that together it should satisfy
 \be
{\tilde k}_{R, j} \propto {\tilde h}_{0j} e^{i \Phi_{R}}~~~\text{and}~~ {\alpha_{R, m}} \propto \text{F}_{\times m}^{DP} \, .\label{eq:filter21}
\ee
The Frobeinus norm of $\boldsymbol{\tilde \K}_R$ is 
$\| \F'^{DP}_\times\|$,  gives the components of 
normalized network cross filter $\boldsymbol{\tilde \K}_R$,
{\be
\tilde k_{R,j} = \tilde h_{0 j} e^{i \Phi_R}~, \hspace{0.2in} \alpha_{R,m} =  \frac{\text{F}^{DP}_{\times m}} { \| \F'^{DP}_{\times} \|}~. \label{eq:filter22}
\ee}
with the corresponding SNR as given by,
{\be
4 \Re \ll[  \sum_{j=1}^N \sum_{m=1}^I \tilde {\tilde \S}_{jm}  ~\tilde \K_{R, jm}  \rr] = \boldsymbol{\rho}_R~. \label{eq:snr2}
\ee}
\end{enumerate}

In summary, the synthetic streams constructed from the over-whitened data streams which captures the individual polarizations in the DP frame are 
\be \tilde {\tilde \z}_L =\sum_{m=1}^I \frac{{\text{F}}^{DP}_{+m}}{\| \F'^{DP}_+\|} \tilde {\tilde X}_{jm},
~~~ \tilde {\tilde \z}_{R} =\sum_{m=1}^I \frac{{\text{F}}^{DP}_{{\times}m}}{\| \F'^{DP}_{\times}\|} \tilde {\tilde X}_{jm}~ , \label{eq:z1z2}
\ee
They together give the total network SNR as the sum squares of individual SNRs. The total signal power in the individual detectors of the network is now distributed among the synthetic streams $\tilde \z_L$ and
$\tilde \z_{R}$, which when processed with filters $\tilde \k_L$ and $\tilde \k_{R}$ independently captures the two polarizations in DP frame. 
By using  Eq.\eqref{eq:filter12}, Eq.\eqref{eq:snr1}, Eq.\eqref{eq:filter22} and Eq.\eqref{eq:snr2}, we can write the respective SNRs as,
{\be
\boldsymbol{\rho}_L = \ll. \langle  \z_L|  \k_L \rangle \rr|_{\n=0}, \hspace{0.5in}\boldsymbol{\rho}_R = \ll. \langle  \z_R|  \k_R \rangle \rr|_{\n=0}.\label{eq:snr3}
\ee}
Thus, we have shown that extending the concept of the matched filter to the network gives us two effective 
synthetic data streams which can be further processed.

\subsection{ Special Case: Same Noise for all Detectors}
In this subsection, we consider an idealistic situation where all the detectors
have same noise PSD, {\it i.e.} $g_m$'s becomes  equal to a constant '$g$' for all detectors. Then from 
Eq.\eqref{eq:z1z2}, we can see that the 'noise free' synthetic streams, $\tilde \z^s_{L,R}$ are nothing but
the projections of network signal matrix $\boldsymbol{\tilde \S}$ on the orthonormal vectors $\hat \F^{DP}_{+,\times}$
 with an over all weight $1/g$. If we expand $\boldsymbol{\tilde \S}$ as in Eq.\eqref{eq:sij}, we can further simplify $\tilde \z^s_{L,R}$ 
into  linear combination of GW polarizations, $\tilde \h_{+,\times}$ similar to a pair of ordinary interferometric detector signals as 
shown below.
\bea
 \tilde \z^s_{L} &=& \frac{\| \F^{DP}_+ \|}{g} \ll( \tilde \h_+ ~ \cos 2\chi ~-~\tilde \h_{\times} ~ \sin 2\chi \rr) \, , \nonumber \\
 \tilde \z^s_{R} &=& \frac{\| \F^{DP}_\times \|}{g} \ll( \tilde \h_+ ~ \sin 2\chi ~+~ \tilde \h_{\times} ~ \cos 2\chi \rr) \, .
\eea
Please note the equivalent antenna pattern of $\tilde \z^s_{R}$ is $\pi/4$ out of phase with that of $\tilde \z^s_{L}$.

\section{Multi-detector Maximum Likelihood Ratio and network Synthetic Streams}
\label{sec:LR}
In this section, we carry out MLR analysis for the inspiral detection with a multi-detector network.
This has been already done in the literatures \cite{apaiprd01,apaiprd06,harryprd11} in different contexts as well as notations.
Here, we construct MLR statistic in much more straightforward way and latter we make connection to the previous works and thus bring all earlier multi-detector inspiral related search formalisms under the same notations.

In multi-detector network MLR  detection technique,the network LLR is maximized over signal
parameters and a test statistic is obtained, which is then compared with the threshold for the detection. 
For high SNR cases, MLR technique is known to be optimal.

\subsection{Network Likelihood Ratio}
Assuming the Gaussian, additive noise in each detector data, the LLR for a multi-detector network with $I$ constituent
detectors is the sum of LLR's of individual antennas and is given by \cite{harryprd11},
\be
\varLambda = \sum_{m=1}^I \langle \x_m| \s_m \rangle -\frac{1}{2} \langle  \s_m| \s_m \rangle~. \label{eq:gamma1}
\ee

Re-arranging terms and little algebra as given in Appendix-\ref{app:LR}, we can re-express the above equation in terms of the
synthetic data streams as
{\small \be 
2 \varLambda = \ll[2 \boldsymbol{\rho_L} \langle  \z_L|   \h_0 e^{i  \Phi_L} \rangle - \boldsymbol{\rho_L}^2 \rr] +\ll[ 2 \boldsymbol{\rho_R} \langle  \z_R|  \h_0 e^{i  \Phi_R} \rangle  - \boldsymbol{\rho_R}^2 \rr] ~. \label{eq:gamma2}
\ee}

We note that Eq.\eqref{eq:gamma2} can be viewed as the sum of LLR of 2 independent synthetic detectors, where $\Phi_{L,R}$ carries the constant phase which incorporates
the initial phase plus the polarization angles and $\boldsymbol{\rho}_{L,R}$ are the SNRs of the two synthetic data streams.

We note that in terms of synthetic streams, the four extrinsic parameters $\ll(\text{A}_0, \phi_a, \epsilon, \Psi \rr)$ are now mapped in to a set of  two effective 
SNRs and 2 effective phases, namely $\ll(\boldsymbol{\rho}_L,\boldsymbol{\rho}_R,  \Phi_L,  \Phi_R \rr)$.

\subsection{Maximization of Network LLR}
Now we maximize $\varLambda$ over the new extrinsic parameters $\ll(\boldsymbol{\rho}_L, \boldsymbol{\rho}_R, \Phi_L,  \Phi_R \rr)$ to obtain MLR
$\hat \varLambda$ \cite{apaiprd01,apaiprd06,harryprd11}. The new extrinsic parameters give the re-parametrized physical parameters $\ll(\text{A}_0, \phi_a, \epsilon, \Psi\rr)$ where the relation between them is summarized in Appendix-\ref{app:extrinsic}.
Below, we maximize $\varLambda$ over the new set; first over the amplitudes and then over the phase respectively.

\bn
\item {\bf Amplitude Maximization} \\
Maximization over $\boldsymbol{\rho}_{L,R}$ is same as that in case of single detector\cite{sanjeev1991} and it results in,
\be
2 \hat{\varLambda}~|_{\boldsymbol{\rho}_L,\boldsymbol{\rho}_R} = \langle  \z_L|  \h_0 e^{i  \Phi_L} \rangle^2 + \langle  \z_R|  \h_0 e^{i  \Phi_R} \rangle^2  ~, \label{eq:gamma3}
\ee
and the MLR amplitude estimates become
\be
\label{eq:MLRrho} 
\boldsymbol{\hat \rho}_L =  \langle  \z_L|  \h_0 e^{i  \Phi_L} \rangle~, \hspace{0.3in} \boldsymbol{\hat \rho}_R =  \langle  \z_R|  \h_0 e^{i  \Phi_R}\rangle.
\ee
\item {\bf Phase Maximization}\\
Since $\Phi_L$ and $\Phi_R$ are independent, maximization of
LLR over them amounts to individually maximizing each term of the sum in Eq.\eqref{eq:gamma3}. 
Thus the ML estimates
of $\Phi_{L,R}$ are,
{\small\be
\hat {\Phi}_L = arg \ll[\sum_{j=1}^N \tilde {\tilde z}_{L,j}~ \tilde h^*_{0j} \rr]~, \hspace{0.1in} \hat {\Phi}_R = arg \ll[\sum_{j=1}^N \tilde {\tilde z}_{R, j}~ \tilde h^*_{0j} \rr].
\ee}
\en
In summary,
\be
2 \hat \varLambda =  16 \ll[ \ll|\sum_{j=1}^N \tilde  {\tilde z}_{L,j}~ \tilde h_{0j}^*  \rr|^2 + \ll|\sum_{j=1}^N \tilde  {\tilde z}_{R, j}~ \tilde h_{0j}^*  \rr|^2 \rr]~. \label{eq:gammamax1}
\ee
We write one individual term in Eq.\eqref{eq:gammamax1} as follows,  
{ \bea
 &&\ll| \sum_{j=1}^N \tilde {\tilde z}_{L,R~j}~ \tilde h_{0j}^*  \rr|^2  \nonumber \\
&&~~=\ll( \Re\ll[ \sum_{j=1}^N \tilde {\tilde z}_{L,R~j} \tilde h_{0j}^* \rr] \rr)^2 + \ll( \Re\ll[ \sum_{j=1}^N \tilde {\tilde z}_{L,R~j} \tilde h_{\pi/2 j}^* \rr] \rr)^2, \nonumber \\
&&~~=\frac{\langle  \z_{L,R} |  \h_0 \rangle^2 + \langle  \z_{L,R} |  \h_{\pi/2} \rangle^2}{16}. \label{eq:gammamax2}
\eea}

Thus the MLR simplifies to
{\be
2 \hat \varLambda =  \langle  \z_L |  \h_0 \rangle^2 + \langle  \z_L |  \h_{\pi/2} \rangle^2 
+\langle  \z_R |  \h_0 \rangle^2 + \langle  \z_R |  \h_{\pi/2} \rangle^2 . \label{eq:gammamax3}
\ee}
This can be described as quadrature sum of powers in synthetic streams $\tilde \z_L$ and $\tilde \z_R$. This is similar to the
single detector statistic which contains the quadrature sum of powers in a single detector data stream.
From Eq.\eqref{eq:snr3} and Eq.\eqref{eq:gammamax1}, we can see under no noise condition\cite{apaiprd01,apaiprd06,harryprd11}, 
\be 2 \hat \varLambda = \boldsymbol{\rho_L}^2 + \boldsymbol{\rho_R}^2~.\ee

\section{Connection to the existing literature}
\label{literaturecomp}
In the GW multi-detector inspiral search, network MLR statistics maximized over 4 extrinsic 
parameters has been formalized in the literature \cite{apaiprd01,harryprd11}. Though the problem is 
same, the parametrization depends on the way the problem is casted. 
However, the final MLR , maximized over the extrinsic parameters is the same.
In this section, we carry out comparison between various formalisms under the same notations as given here,
which till now is not been done in the literature so far.

\subsection{Synthetic Streams and Harry-Fairhurst \cite{harryprd11}  Approach}
In \cite{harryprd11}, the authors casted the multi-detector MLR problem into
the F-statistic. The polarizations were written down in terms of the linear
combination of the 4 amplitudes on which the extrinsic parameters are mapped {\it i.e.}
\be
(A_0, \phi_a, \epsilon, \Psi) \Rightarrow (\cA_1,\cA_2, \cA_3, \cA_4)~.
\ee
For explicit relations, please visit Appendix-\ref{app:extrinsic}.
Latter they transform wave frame to DP frame so that Maximum Likelihood SNR square, $2 \hat{\varLambda}$ is
simplified to quadrature sum of powers in $+$ and $\times$ polarizations. Here, the maximization over the four amplitudes 
is carried out at a time.

Below, we derive the multi-detector MLR of \cite{harryprd11} Eq. (2.33) starting from the notations in this paper.
 
In Eq.\eqref{eq:gamma3},
{\small \bea
\langle  \z_{L,R} |  \h_0 \rangle  &=& \frac{4}{\| \F^{'DP}_{+,\times}\|}  \Re\ll[ \sum_{j=1}^N \sum_{m=1}^I \tilde{\tilde X}_{jm}~ {h_0}_j^* \text{F}_{+,\times m}^{DP}  \rr]  \nonumber \\
&=&  \frac{1}{ \| \F^{'DP}_{+,\times}\|} \sum_{m=1}^I \langle  \x_m| \h_0 \text{F}_{+,\times m}^{DP}\rangle~, \nonumber \\
\text{and}
~~\langle  \z_{L,R} |  \h_{\pi/2} \rangle&=& \frac{1}{ \| \F^{'DP}_{+,\times}\|} \sum_{m=1}^I \langle  \x_m | \h_{\pi/2} \text{F}_{+,\times m}^{DP}\rangle.
\eea}
Further,
 {\small \be 
\| \F^{'DP}_{+,\times}\|^2 = \sum_{m=1}^I g_m^2  \text{F}_{+,\times m}^2 =  \sum_{m=1}^I \langle \h_0 \text{F}_{+,\times m}| \h_0 \text{F}_{+,\times m}\rangle~. \nonumber
\ee}

Now substituting back in Eq.\eqref{eq:gammamax3},
{\small \bea 
2 \hat \varLambda &=& \frac{\ll[\sum_m \langle \x_m |\h_0 \text{F}_{+ m}^{DP} \rangle \rr]^2 +\ll[\sum_m \langle \x_m |\h_0 \text{F}_{+ m}^{DP} \rangle \rr]^2 }{\sum_m \langle \h_0 \text{F}_{+ m}| \h_0 \text{F}_{+ m} \rangle} \nonumber \\
&+& \frac{\ll[\sum_m  \langle \x_m |\h_0 \text{F}_{\times m}^{DP} \rangle \rr]^2 +\ll[\sum_m \langle \x_m |\h_0 \text{F}_{\times m}^{DP} \rangle \rr]^2 }{\sum_m \langle \h_0 \text{F}_{\times m}| \h_0 \text{F}_{\times m} \rangle }~. \label{eq:gammamax4}
\eea}
Absorbing the summation over $m$ following the definition Eq.(2.21) of \cite{harryprd11}, Eq.\eqref{eq:gammamax4} would go to Eq.(2.33) of \cite{harryprd11}
\footnote{$\langle \mathbf{a}| \mathbf{b}\rangle$ is same as $(\mathbf{a}| \mathbf{b} )$ defined in Eq.(2.17) of \cite{harryprd11}.}.
Also one  can show that $\tilde{\tilde \z}_{L,R}$ are related to over-whitened synthetic streams $o_{+,\times}$ defined in Eq.(2.35) of \cite{harryprd11} as follows,
\be
o_{+,\times}=\| \F^{'DP}_{+,\times} \| ~\tilde{\tilde \z}_{L,R}. 
\ee
The two pairs of synthetic streams differ by  constants which is different for both the synthetic streams. The final multi-detector MLR matches to Eq. \eqref{eq:gammamax3} as expected.

\subsection{Synthetic Streams and Pai et al \cite{apaiprd01} Approach}
In \cite{apaiprd01}, the multi-detector coherent statistic was obtained by successive maximization of amplitude $A_0$, initial
phase $\phi_a$ similar to single detector statistic. The polarization angles $(\epsilon,\psi)$ are maximized at a time
using the symmetry properties of the rotation group and Gel-Fand functions. The MLR thus obtained contains the
sum square of four terms as is shown in Eq.(4.11) of \cite{apaiprd01} similar to Eq.(2.33) of \cite{harryprd11} and Eq.\eqref{eq:gammamax3} above. We explicitly
give the Eq.(4.11) of \cite{apaiprd01} below.
\bea \label{SSnew}
2 \hat \varLambda &=& {|\hat{\sf v}^+ \cdot {\sf C}|}^{2} 
+{|\hat{\sf v}^+ \cdot {\sf C}|}^{2} \\
 &=& (c_0^+)^2 + (c_{\pi/2}^+)^2 + (c_0^-)^2 + (c_{\pi/2}^-)^2 \nonumber 
\eea
where the elements of $I$-dimensional complex vector ${\sf C}$,
\be \label{Cstar}
C^m = c_0^m + i c_{\pi/2}^m 
\ee
combines the correlations of the two quadratures of the normalized template with the data 
with $c_0^m = \langle \h_0^m|\x^{m}\rangle_{(I)}$ and $c_{\pi/2}^m = \langle \h_{\pi/2}^m|\x^{m} \rangle_{(m)}$.
Further,
\be
{\hat {\sf v}}^\pm =\frac{{\sf v}^{\pm}}{\|{\sf v}^{\pm}\|} = \frac{\widehat{\bf \Re(d)} \pm {\widehat {\bf \Im(d)}}}{\parallel  ({\widehat {\bf \Re(d)}} \pm {\widehat {\bf \Im(d)}}) \parallel}
\ee
is a pair of real unit vectors which span the 2-dimensional polarization plane in the $I$-dimensional space. Thus, if we take
a representative individual term in Eq.\eqref{SSnew}, it becomes
\be \label{term1}
(c_0^+)^2 = \ll(\sum_{I=m}^I \frac{{\sf v}^+_m}{\parallel {\sf v} \parallel} \langle \x_m, \h_0 \rangle \rr)^2 \equiv \langle \z_{+}|\h_0 \rangle^2
\ee
Thus, based on Eq.\eqref{term1} the corresponding synthetic streams are
\be \tilde {\tilde \z}_{+j} \equiv \sum_{m=1}^I \frac{{\sf{v}}^+_{m}}{g_m \| \sf{v}^+\| } \tilde {\tilde X}_{jm},
~~~~\tilde {\tilde \z}_{-j} \equiv \sum_{m=1}^I \frac{{\sf{v}}^-_{m}}{g_m \| \sf{v}^-\| } \tilde {\tilde X}_{jm}.
\label{eq:z1z2p}
\ee
which give the SNR's $\rho_+$ and $\rho_-$ such that in no noise case,
\be
2 \hat{\Lambda} = \rho_+^2 + \rho_-^2.
\ee

\subsection{SNRs in two pairs of Synthetic Data Streams}
In this section, we explicitly show that the two data streams namely; 
$\tilde \z_{L,R}$ and $\tilde \z_{+,-}$ are not same though they give the same network SNR
when combined in pairs. 

Following Eq.\eqref{eq:z1z2} and Eq.\eqref{eq:z1z2p}, the two pairs of synthetic streams are constructed using a pair of two real vectors namely; 
$\{\F'^{DP}_{+,\times}\}$ and $\{\sf{v}_\pm\}$ respectively. We write down these
real vectors explicitly in terms of $\d$ as,
\bea
\F'^{DP}_+ &=& \Re[\d e^{-i\frac{\delta}{2}}]~,                   \hspace{0.5in} \F'^{DP}_\times = \Im[\d e^{-i\frac{\delta}{2}}]~,   \\
{\sf{v}}_+ &=& \Re[\d e^{-i \alpha}]~,                   \hspace{0.67in} {\sf{v}}_- = \Re[\d e^{i \alpha}]~,
\eea
with $\alpha =  {\frac{1}{2}\rm cos^{-1}\ll[-\frac{|\d^T \d | \cos \delta}{\d^H \d} \rr]}$ and $\delta =\arg(\d^T\d)$ as defined earlier.

\begin{figure*}
\centering
\subfloat[LHV]{\includegraphics[width=0.6\textwidth]{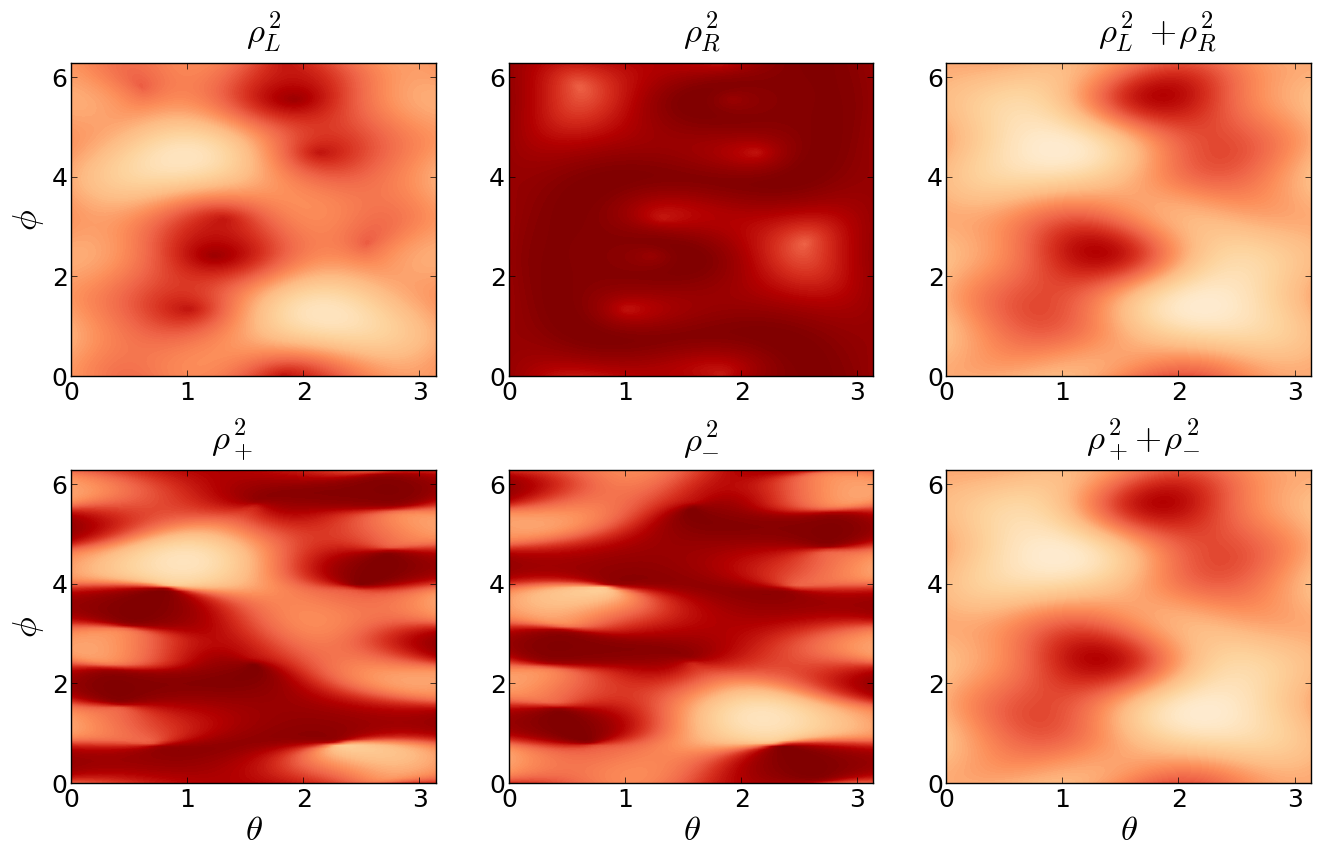}}\vspace{-0.3cm}\\
\subfloat[LVI]{\includegraphics[width=0.6\textwidth]{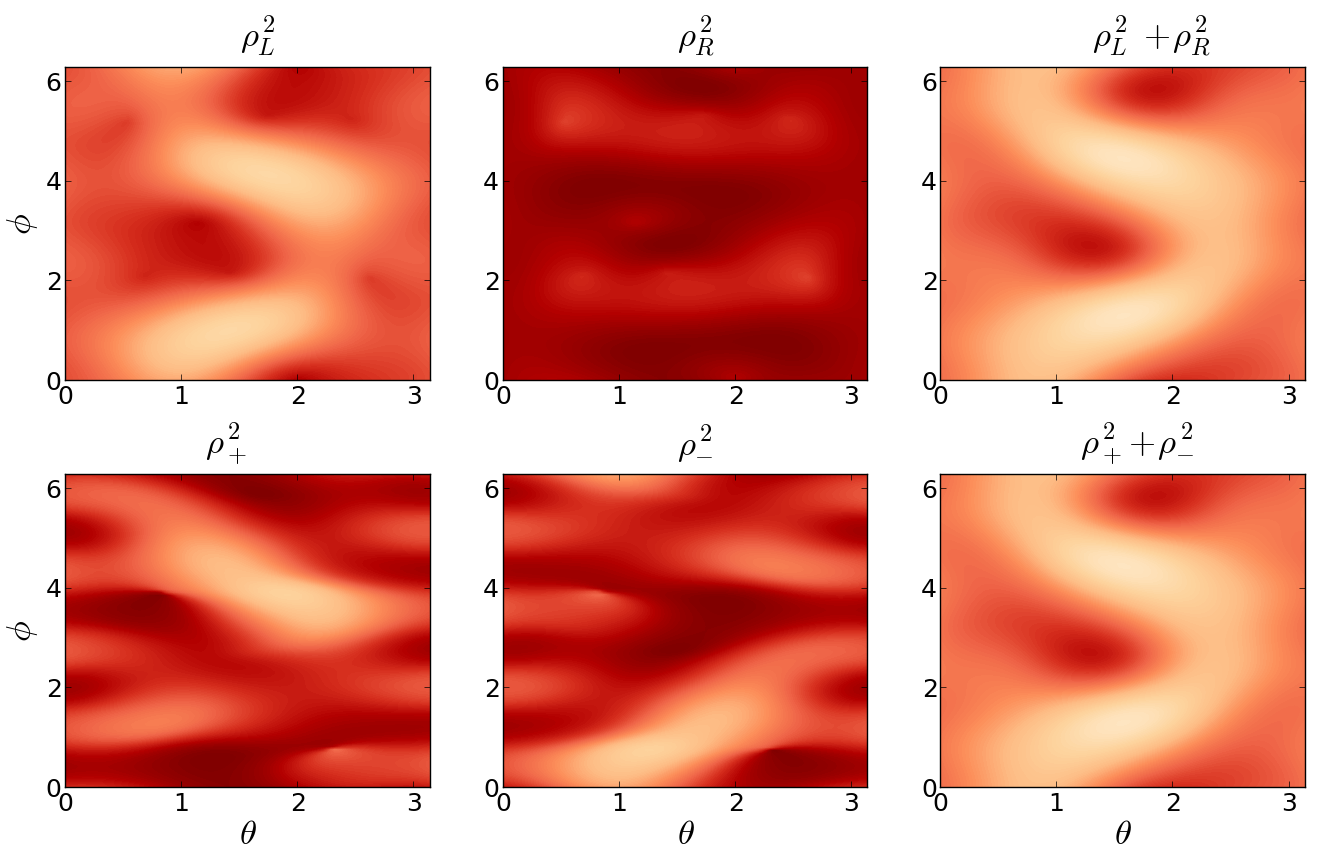}}\vspace{-0.3cm}\\
\subfloat[LHI]{\includegraphics[width=0.6\textwidth]{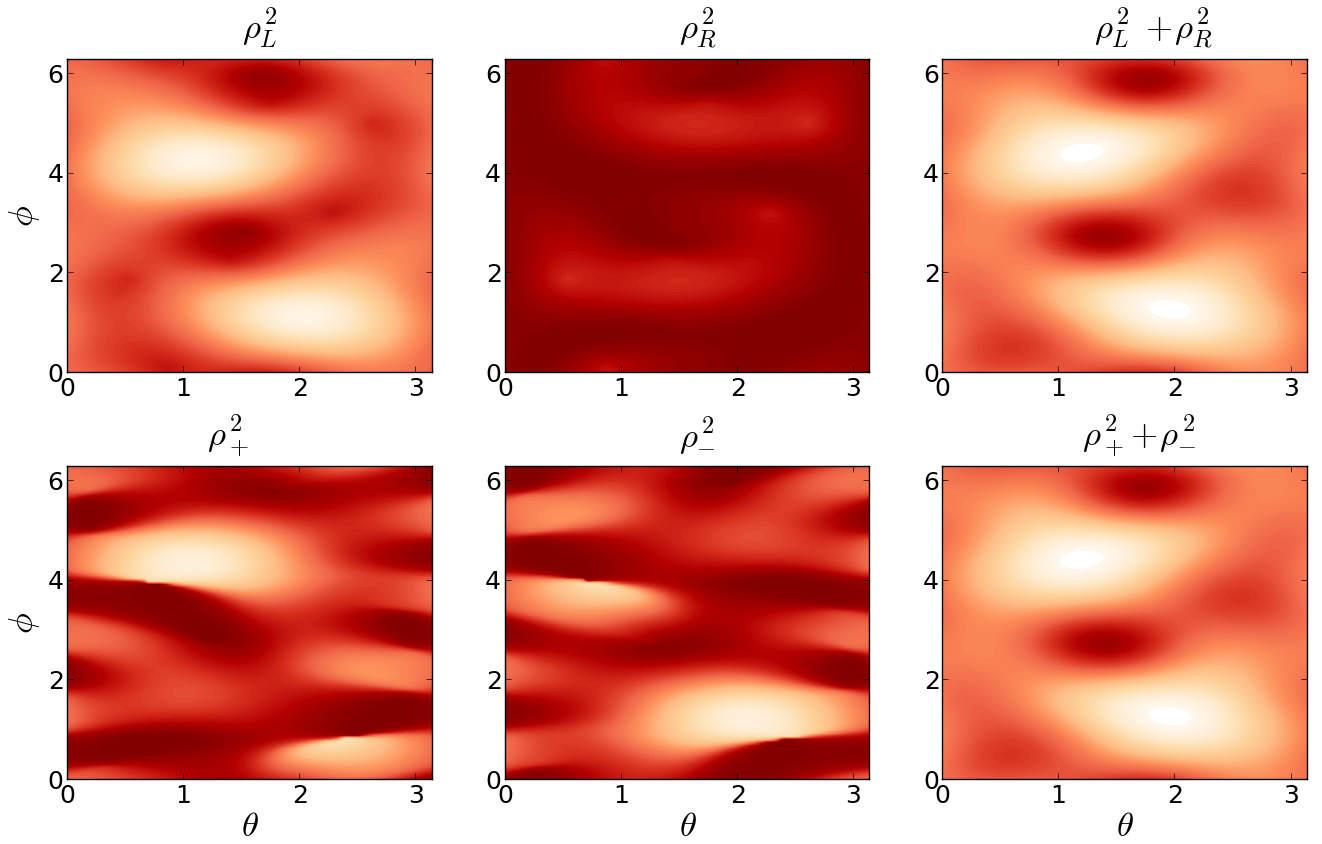}}\vspace{-0.3cm}\\
\subfloat{\includegraphics[width=0.6\textwidth]{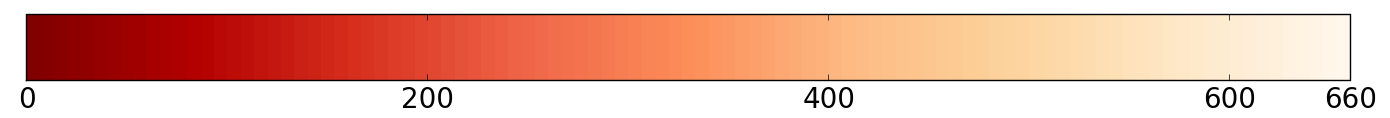}}
\caption{\label{fig1} Directional SNR Squares  $\boldsymbol{\rho^2_L}$ , $\boldsymbol{\rho^2_{R}}$, $\boldsymbol{\rho^2_+}$ , $\boldsymbol{\rho^2_{-}}$ and $\boldsymbol{\rho^2}_{net}$ for various three network configurations with $m_1,m_2=1.4,~$ \\ $\epsilon=\pi/4,~\Psi=\pi/4,~ r= 150MPc.$ with same noise spectral densities for all detectors.}
\end{figure*}

\begin{figure*}
\centering
\subfloat[LHVK]{\includegraphics[width=0.6\textwidth]{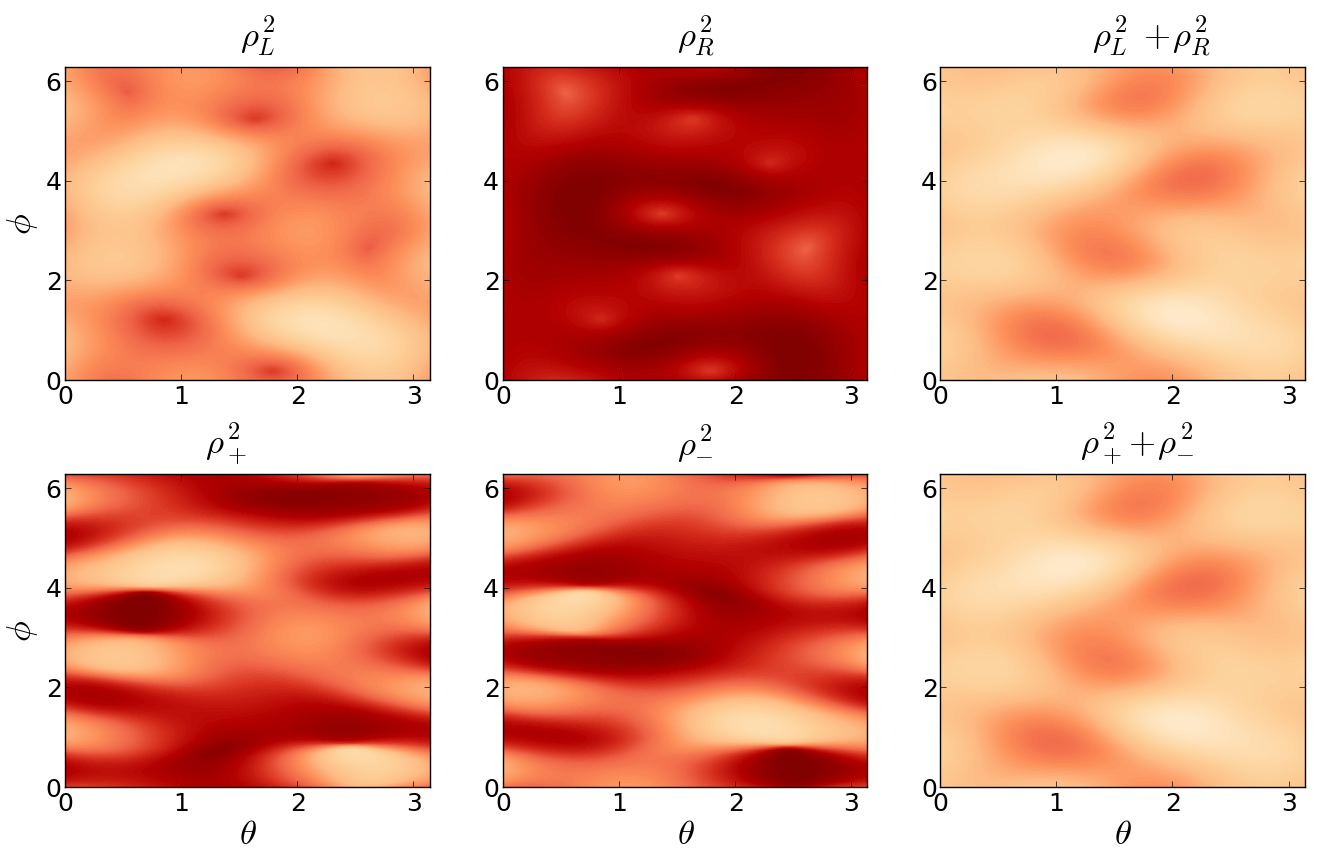}}\vspace{-0.3cm}\\
\subfloat[LHVI]{\includegraphics[width=0.6\textwidth]{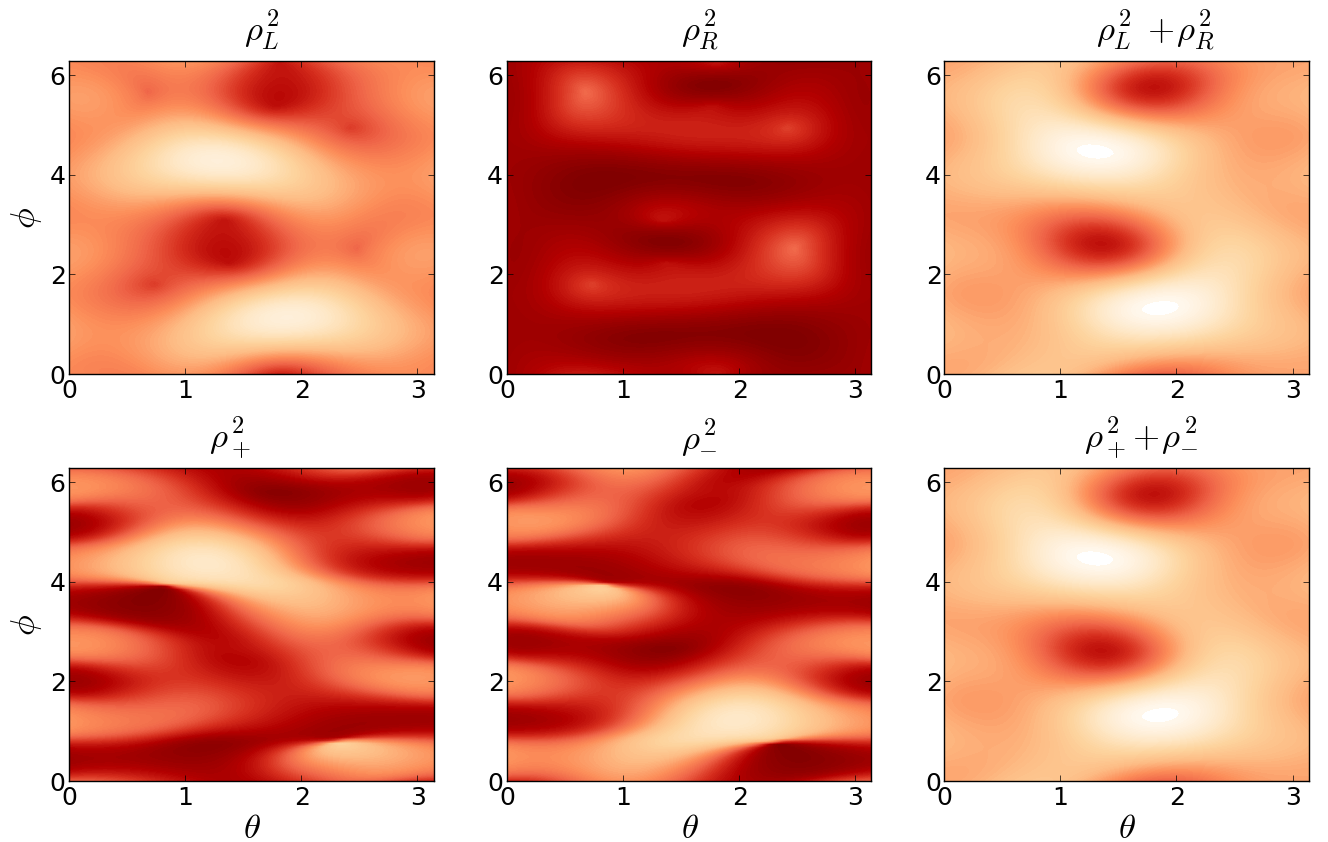}}\vspace{-0.3cm}\\
\subfloat[LVKI]{\includegraphics[width=0.6\textwidth]{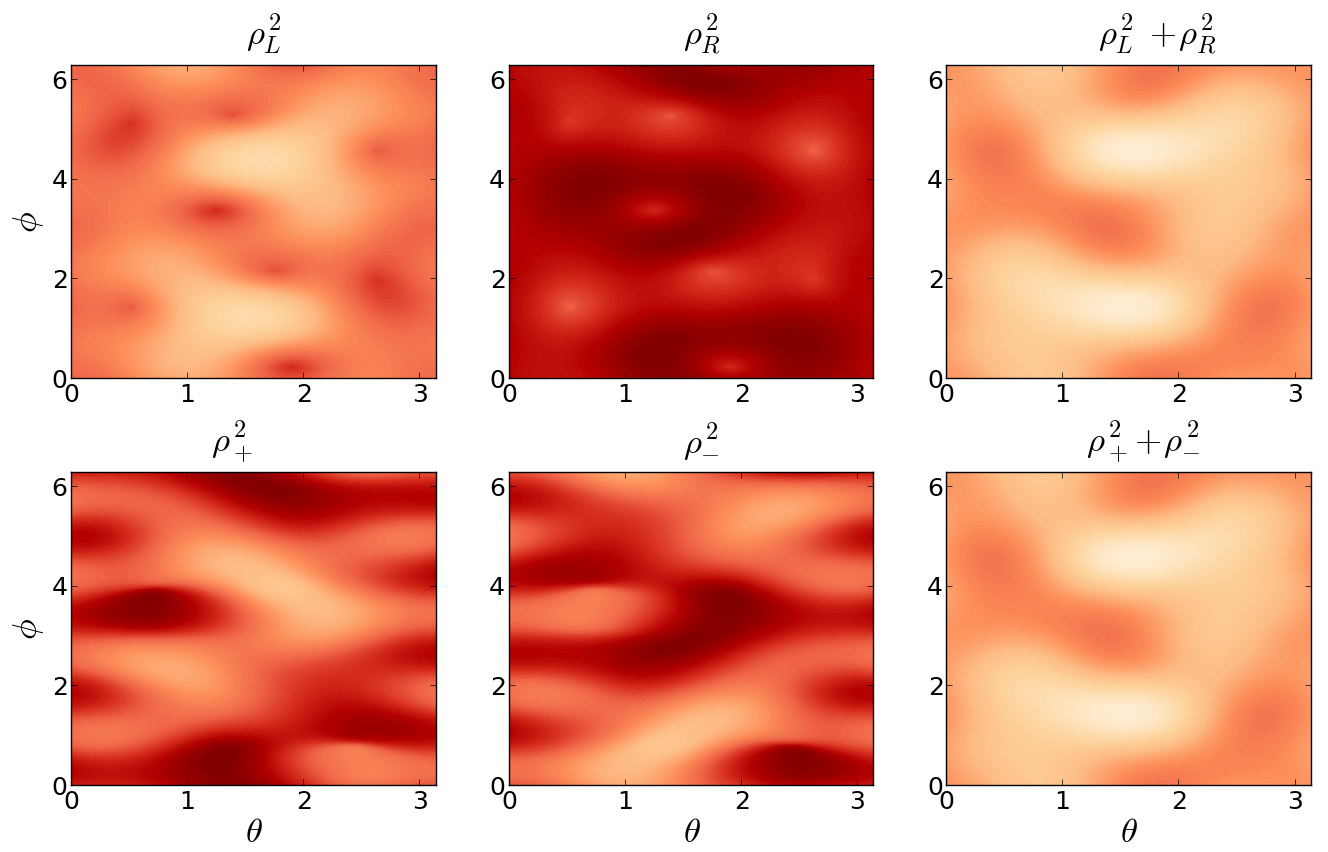}}\vspace{-0.3cm}\\
\subfloat{\includegraphics[width=0.6\textwidth]{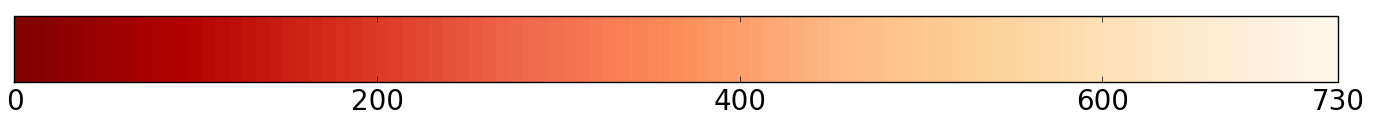}}
\caption{\label{fig2} Directional SNR Squares  $\boldsymbol{\rho^2_L}$ , $\boldsymbol{\rho^2_{R}}$, $\boldsymbol{\rho^2_+}$ , $\boldsymbol{\rho^2_{-}}$ and $\boldsymbol{\rho^2}_{net}$ for various four detector network configurations with ~$m_1,m_2=1.4,~$ \\$\epsilon=\pi/4,~ \Psi=\pi/4,~ r= 150MPc.$ with same noise spectral densities for all detectors.}
\end{figure*}

\begin{figure*}
\centering
\subfloat{\includegraphics[width=0.6\textwidth]{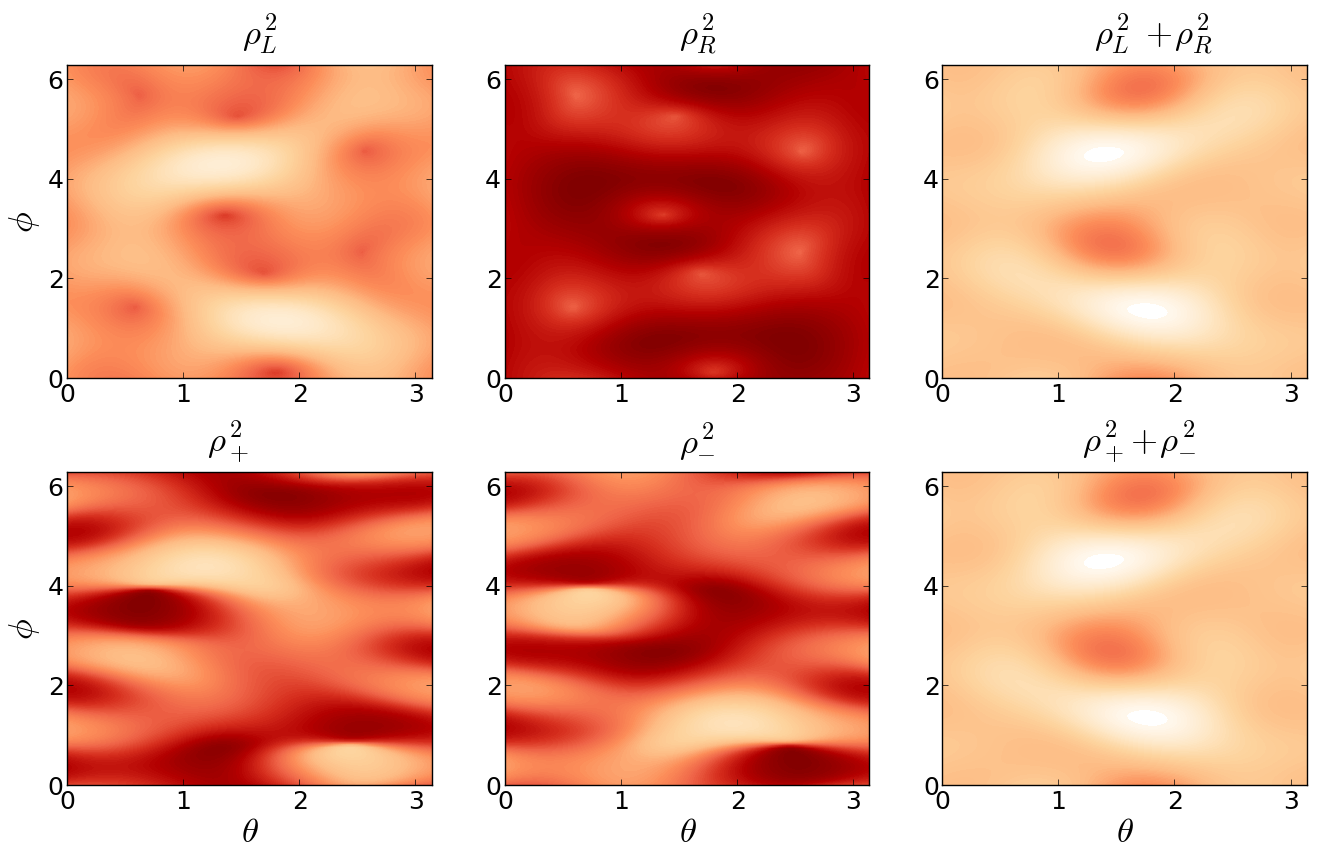}}\vspace{-0.3cm}\\
\subfloat{\includegraphics[width=0.6\textwidth]{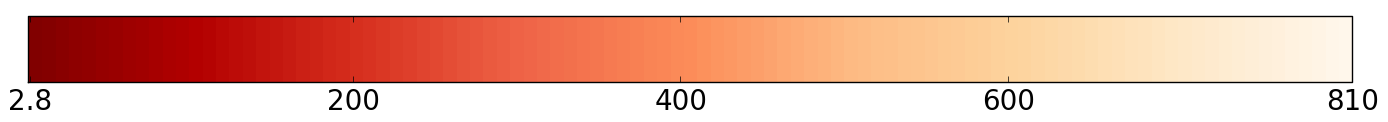}}
\caption{\label{fig3} Directional SNR Squares  $\boldsymbol{\rho^2_L}$ , $\boldsymbol{\rho^2_{R}}$, $\boldsymbol{\rho^2_+}$ , $\boldsymbol{\rho^2_{-}}$ and $\boldsymbol{\rho^2}_{net}$ for network LHVKI with $m_1,m_2=1.4,~ \epsilon=\pi/4,~ \Psi=\pi/4,~$ \\$r= 150MPc.$ with same noise spectral densities for all detectors.}
\end{figure*}

The figures, Fig. \ref{fig1}, \ref{fig2}, \ref{fig3} are the contour plots of SNR squares of the individual synthetic data streams
as well as the network SNR square as a function of the source location for a given polarization in various combinations of 3 detector networks,
4 detector networks and 5 detector network respectively. We assume all the detectors with "zero-detuning, high power" Advanced LIGO noise curve
given by Eq.(4.7) of \cite{AdvLigoPSD}. We took the binary system with component masses of $(1.4,1.4) M_{\odot}$ located at the distance of 150 Mpc
as source with $\epsilon = \psi = \pi/4$. We consider networks constructed out of LIGO-Livingston(L), LIGO-Hanford(H), Virgo(V), proposed KAGRA(K)
and detector in India denoted by (I)\footnote{Hypothetically we take Pune, India as the location for the detector in India.}. 

Several distinct features are to be noted in the SNR figures of the two pairs; namely there is definite symmetry in the 
SNR pattern $\rho_{L,R}^2$ whereas $\rho_{\pm}^2$ lacks that feature. Further, on average network SNR seemed to have distributed between the $\rho_+$ and 
$\rho_-$ equally whereas $\rho_L$ carries large fraction of the network SNR. Some of these features can be used to construct consistency
test for the targeted directional search which is under consideration. 

\section{Conclusion}
The multi-detector interferometric GW network can be described as a pair {\it effective} multi-detector antennas which captures most
of the features of many detectors acting in phase coherent fashion. Till now, in the compact binary coalescence literature, the two synthetic
stream pertaining to the two polarizations were always a by-product of the MLR analysis of the multi-detector analysis. In this work, for the first time,
the authors have derived the synthetic data streams using the matching filtering idea applied to the network combined with the singular-value-decomposition
technique applied to the network SNR vector.

Then, the network LLR naturally emerges as the sum of the LLR of the single synthetic stream LLR. The MLR over the new parameters
namely the two amplitudes and the two phases is a straightforward task. We further, demonstrate that the dominant polarization plane naturally
emerges out of the SVD of the SNR vector. 

Connecting this work to the existing literature; namely \cite{apaiprd01} and \cite{harryprd11}, we explicitly show that the two synthetic streams discussed in the
earlier works are distinct and they can be related through the network constructs. Though this work is theoretical in nature, it combines all the
existing formalisms of the multi-detector pertaining to the compact binary coalescence. We are further investigating the properties of these streams
and its possible applications in the inspiraling binary search namely to develop the consistency tests as well as to carry out efficient all-sky search 
with a global detector network. 

\section{Acknowledgement}
This work is supported by AP's SERC Fast Track Scheme For Young Scientists. HK
thanks Albert Einstein Institute, Hannover for hospitality for stay during where part of the manuscript writing was carried out. 
The authors would like to thank  S. Fairhurst, B. S. Santhyaprakash and Gianluca Guidi for useful discussion and helpful comments on this work.
The main result of this work is presented in the LIGO-Virgo Scientific Meeting at Hannover, 2013 (LIGO laboratory document number: LIGO-G1301109).
This document has been assigned LIGO laboratory document number LIGO-P1300229.


\appendix
\section{Likelihood Ratio}
\label{app:LR}
The network log Likelihood Ratio, 
{ \small
\be 
\varLambda =  \sum_{m=1}^I \langle  \x_m|  \s_m \rangle -\frac{1}{2} \langle  \s_m| \s_m \rangle \nonumber 
\ee}

{\small \be
= \sum_{m=1}^I ~4\Re \ll[ \sum_{j=1}^N \tilde{\tilde \X}_{jm} \tilde \S_{jm}^* \rr] - 2 \ll[\sum_{j=1}^N \frac{\ll| \tilde \S_{jm}\rr|^2}{\N_{jm}} \rr]. \label{appeq:gamma}
\ee }

We use Eq.\eqref{eq:z1z2} to express the network LLR in terms of $(P_L,P_R,\Phi_L,\Phi_R)$ etc, we get
{\small \bea
\sum_{m=1}^I 4\Re \ll[ \sum_{j=1}^N \tilde{\tilde \X}_{jm}~ \tilde S_{jm}^* \rr] 
&=& 4 \text{A}_0 \text{P}_L~ \Re \ll[ \sum_{j=1}^N  \sum_{m=1}^I \tilde{\tilde \X}_{jm} \text{F}^{DP}_{+ m} ~\tilde h^*_{0j} e^{-i \Phi_L} \rr] \nonumber \\
&+& 4 \text{A}_0 \text{P}_R~ \Re \ll[ \sum_{j=1}^N  \sum_{m=1}^I \tilde{\tilde \X}_{jm} \text{F}_{\times m}^{DP} ~\tilde h^*_{0j} e^{-i \Phi_R } \rr] \nonumber
\eea}
{\small \be
= \text{A}_0 \ll[\text{P}_L \| \F^{'DP}_+ \| \langle  \z_L| \h_0 e^{i \Phi_L} \rangle
+  \text{P}_R \| \F^{'DP}_{\times} \| \langle  \z_R| \h_0 e^{i \Phi_R} \rangle \rr]~.
\ee}

\vspace{0.4cm}
Eq.\eqref{eq:snrnet2} and Eq.\eqref{eq:dpant1}, gives
\be
\text{A}_0 \text{P}_L \| \F^{'DP}_{+} \| = \boldsymbol{\rho}_L~,~~~~ \text{A}_0 \text{P}_R \| \F^{'DP}_{\times} \| = \boldsymbol{\rho}_R~.
\ee

Also, from Eq.\eqref{eq:snrnet1} one can easily show that the second terms in 
Eq.\eqref{appeq:gamma} is half of network SNR square, $\boldsymbol{\rho}_L^2 +\boldsymbol{\rho}_R^2$.

Substituting back in Eq.\eqref{appeq:gamma},
{\small \be
2 \varLambda = \ll[2 \boldsymbol{\rho}_L \langle \z_L|  \h_0 e^{i  \Phi_L} \rangle - \boldsymbol{\rho}_L^2 \rr] +\ll[2 \boldsymbol{\rho}_R \langle \z_R| \h_0 e^{i  \Phi_R} \rangle  -\boldsymbol{\rho}_R^2 \rr]. 
\ee}

\section{Relation between old and new extrinsic parameters}
\label{app:extrinsic}
MLR is obtained by maximizing network LLR over the four extrinsic parameters, which are the functions of  physical parameters, $\ll(\text{A}_0, \phi_a, \epsilon, \Psi \rr)$.
As we discussed earlier, the choice of these functions depend  on the formalism. But the final results for various approaches will remain same.

The extrinsic parameters used in this paper are,
{\small \bea
\boldsymbol{\rho}_L &=&  \text{A}_0 \|\F_+^{DP}\| \sqrt{ \ll(\frac{1+\cos^2 \epsilon}{2} \rr)^2 \cos^2 2 \chi + \cos^2 \epsilon \sin^2 2 \chi }, \nonumber \\
\boldsymbol{\rho}_R &=& \text{A}_0 \|\F^{DP}_2\| \sqrt{ \ll(\frac{1+\cos^2 \epsilon}{2} \rr)^2 \sin^2 2 \chi + \cos^2 \epsilon \cos^2 2 \chi }, \nonumber \\
\Phi_L &=& {\rm \tan^{-1}}\ll[{\rm tan}(2 \chi) ~~ \frac{2\cos \epsilon}{1 + \cos^2 \epsilon} \rr] + \phi_a, \nonumber \\
\Phi_R &=& {\rm \tan^{-1}}\ll[{\rm -cot}(2 \chi) ~~ \frac{2\cos \epsilon}{1 + \cos^2 \epsilon} \rr] + \phi_a.
\eea}
In \cite{harryprd11}, the maximization of LLR is done over a set of derived amplitude parameters, 
{\small
\bea
\cA_1 &=& \text{A}_0 \ll[ \frac{1+\cos^2 \epsilon}{2} \cos \phi_a \cos 2 \Psi  - \cos \epsilon \sin \phi_a \sin 2 \Psi \rr]~, \nonumber \\
\cA_2 &=&  \text{A}_0\ll[ \frac{1+\cos^2 \epsilon}{2} \cos \phi_a \sin 2 \Psi  + \cos \epsilon \sin \phi_a \cos 2 \Psi \rr]~, \nonumber \\
\cA_3 &=& \text{A}_0 \ll[- \frac{1+\cos^2 \epsilon}{2} \sin \phi_a \cos 2 \Psi  - \cos \epsilon \cos \phi_a \sin 2 \Psi \rr]~, \nonumber \\
\cA_4 &=& \text{A}_0 \ll[- \frac{1+\cos^2 \epsilon}{2} \sin \phi_a \sin 2 \Psi  + \cos \epsilon \cos \phi_a \cos 2 \Psi \rr]~. \nonumber \\ \label{eq.extrinsic1}
\eea}
These are related to $\ll(\boldsymbol{\rho}_L,\boldsymbol{\rho}_R, \Phi_L,  \Phi_R \rr)$ as follows, 
{\small
\bea
\boldsymbol{\rho}_L &=& \|\F'^{DP}_+ \| \ll| (\cA_1 -i \cA_3) \cos \frac{\delta}{2} ~+~(\cA_2 -i \cA_4) \sin \frac{\delta}{2} \rr|~, \nonumber \\
\boldsymbol{\rho}_R &=& \|\F'^{DP}_\times \| \ll| (\cA_2 -i \cA_4) \cos \frac{\delta}{2} ~-~ (\cA_1 -i \cA_3) \sin \frac{\delta}{2} \rr|~, \nonumber \\
\Phi_L &=& arg \ll[ (\cA_1 -i \cA_3) \cos \frac{\delta}{2} ~+~(\cA_2 -i \cA_4) \sin \frac{\delta}{2} \rr]~, \nonumber \\
\Phi_R &=& arg \ll[ (\cA_2 -i \cA_4) \cos \frac{\delta}{2} ~-~ (\cA_1 -i \cA_3) \sin \frac{\delta}{2} \rr]. 
\eea}

\section{Likelihood Estimates of Polarization angles}
\label{sec:LREst}
As we discussed earlier, a network of detectors can recover the polarization information of GW.
Since $\z_L$ and $ \z_R$ are the equivalent detectors of the network, we can obtain 
the estimates of polarization angles $\hat \epsilon$ and $\hat \Psi$ in terms of their $SNR$s.

By using Eq.\eqref{eq:snrnet2},Eq.\eqref{eq:dpant1} and definition of $ \Phi_{L,R}$ we define,

{\small \bea 
Y &\equiv& \frac{\boldsymbol{\rho}_L \| \F^{DP}_\times\|\e^{i  \Phi_L}}{\boldsymbol{\rho}_R \| \F^{DP}_+\|\e^{i  \Phi_R }} = \frac{\cos 2 \chi ~\frac{1+\cos^2 \epsilon}{2} + i  \sin 2 \chi~ \cos \epsilon} {\sin 2 \chi ~\frac{1+\cos^2 \epsilon}{2} - i  \cos 2 \chi~ \cos \epsilon}
 \nonumber \\
&=& i \frac{T^*_{2+2}(\chi,\epsilon,0)+ T^*_{2-2}(\chi,\epsilon,0)}{T^*_{2+2}(\chi,\epsilon,0)- T^*_{2-2}(\chi,\epsilon,0)}~, \label{eq:yest}
\eea} 

This implies,
\be
\frac{ T_{2-2}(\chi,\epsilon,0)}{T_{2+2}(\chi,\epsilon,0)} = \frac{i Y^* - 1 }{i Y^*+1}~. 
\ee 

Then polarization angles can be expressed in terms of $Y$ as follows.
{\small \bea
\cos 4  \Psi &=&  \cos \ll[ arg\ll( \frac{ T_{2-2}}{T_{2+2}} \rr)+ \delta \rr] \nonumber \\
&=&  \cos \ll[ arg\ll(\frac{i Y^*-1 }{i Y^*+1 } \rr)+ \delta \rr], \label{eq:psiest}
\eea}
and
{\be 
\cos  \epsilon ~=~ \frac{1 - \sqrt{\ll| \frac{ T_{2-2}}{T_{2+2}}\rr|}}{1 + \sqrt{\ll| \frac{ T_{2-2}}{T_{2+2}}\rr|}} 
~=~  \frac{1 - \sqrt{\ll| \frac{i Y^*-1 }{i Y^*+1 } \rr|}}{1 + \sqrt{\ll| \frac{i Y^*-1 }{i Y^*+1 }  \rr|}}~. \label{eq:epsiest}
\ee}
However, when the noise is present $\boldsymbol{\rho}_L, \boldsymbol{\rho}_R, \Phi_L, \Phi_R$ are estimated using  MLR approach.
Then using Eq.\eqref{eq:yest}, $Y$ is constructed out of  $\boldsymbol{\hat \rho}_L, \boldsymbol{\hat \rho}_R, \hat \Phi_L, \hat \Phi_R$ estimates. Thus $(\hat \epsilon,\hat \Psi)$ 
is obtained by Eq.\eqref{eq:psiest} and Eq.\eqref{eq:epsiest}, with newly constructed $Y$.

\bibliography{paper}
\end{document}